\documentclass[]{spie}  

\usepackage[table]{xcolor}
\usepackage{amssymb}
\usepackage{pifont}
\usepackage{eso-pic}

\usepackage{soul}
\usepackage{xcolor}

\usepackage{amsmath,amsfonts,amssymb}
\usepackage{subcaption}
\usepackage{graphicx}
\usepackage{cite} 
\usepackage{times}
\usepackage{epsfig}
\usepackage{amsmath}
\usepackage{nccmath}
\usepackage{amssymb}
\usepackage{mwe}
\usepackage{acro}
\usepackage{amssymb}
\usepackage{xcolor,colortbl}
\usepackage{tabularx}
\usepackage{relsize}
\usepackage{pifont}
\usepackage{booktabs} 
\usepackage{multirow}
\usepackage{multicol}
\usepackage{adjustbox}
\usepackage{float}
\usepackage{graphicx}
\usepackage{makecell}
\usepackage{tabu}
\usepackage[colorlinks=true, allcolors=blue]{hyperref}
\usepackage[capitalize]{cleveref}
\usepackage[
  paperwidth=8.5in,
  paperheight=11in,
  textwidth=6.75in,
  textheight=8.75in,
  centering
]{geometry}
\usepackage{eso-pic}

\title{Interpretable Multiple-Instance Learning of Glomerular Pathomics and Subtype Ratios for eGFR Classification}

\author[a]{Tianyu Shi}
\author[b]{Yuechen Yang}
\author[c]{Leiyue Zhao}
\author[d]{Daniel Reisenbüchler}
\author[e]{Jing Jin}
\author[f]{Thangamani Muthukumar}
\author[f]{Kenji Ikemura}
\author[f,g]{Mert R. Sabuncu}
\author[f,h]{Yihe Yang}
\author[f]{Ruining Deng}

\affil[a]{Sichuan University, Chengdu, 610065, CN}
\affil[b]{Vanderbilt University, Nashville, TN, 37235, USA}
\affil[c]{Johns Hopkins University, Baltimore, MD 21218, USA}
\affil[d]{University of Regensburg, Regensburg, Bavaria 93053, DE}
\affil[e]{Northwestern University, Evanston, IL 60208, USA}
\affil[f]{Weill Medical College of Cornell University, New York, NY 10065, USA}
\affil[g]{Cornell Tech, New York, NY 10044, USA}
\affil[h]{Northwell Health, New Hyde Park, NY 11040, USA}

\authorinfo{Corresponding author: Ruining Deng\\
E-mail: rud4004@med.cornell.edu}

\begin{document} 

\maketitle

\begin{abstract}
Chronic kidney disease is characterized by structural alterations in renal tissue and progressive loss of kidney function. Kidney-biopsy whole-slide images (WSIs) provide important histological evidence for evaluating renal injury, with glomerular changes serving as informative indicators of disease severity. Glomerular subtype ratios are established quantitative measures that provide a compact and clinically meaningful summary of glomerular composition. Beyond these categorical measurements, individual glomeruli also contain quantitative variations in morphology, staining intensity, and texture that may reflect additional aspects of renal injury. However, representing pathomic information using WSI-level mean values may reduce glomerulus-level heterogeneity and diminish the contribution of less prevalent but informative patterns. In this proof-of-concept study, we investigated whether modeling individual glomeruli using multiple-instance learning (MIL) could provide information complementary to conventional subtype ratios in the application of estimated glomerular filtration rate (eGFR) classification. We analyzed 162 kidney-biopsy WSIs from 162 patients and categorized renal function as $\mathrm{eGFR}<60$ or $\mathrm{eGFR}\geq60~\mathrm{mL/min/1.73\,m^2}$. Quantitative size, shape, intensity, and texture features were extracted from segmented glomeruli. Each patient was represented as a bag of glomerular instances, and an attention-based MIL model learned to aggregate their heterogeneous pathomic profiles. We evaluated subtype ratios, pathomics, and their combination using ABMIL, with a multilayer perceptron (MLP) applied to mean-pooled pathomic features as the baseline. In five-fold cross-validation, the subtype-ratio-only model achieved an area under the receiver operating characteristic curve (AUC) of $0.7262 \pm 0.0596$, confirming the utility of conventional ratio-based assessment. MIL using pathomics alone achieved an AUC of $0.7362 \pm 0.0788$, whereas integrating pathomics with subtype ratios produced the highest mean AUC of $0.7630 \pm 0.0665$. SHapley Additive exPlanations (SHAP) were used to examine the contributions of individual quantitative features. These results demonstrate the value of established glomerular subtype ratios and suggest that MIL-based pathomics can complement them by capturing additional glomerulus-level information for quantitative kidney-biopsy analysis.
\end{abstract}


\section{Introduction}  
\label{sec:intro}

\begin{figure}[htbp]
  \centering
  \includegraphics[width=1.0\textwidth]{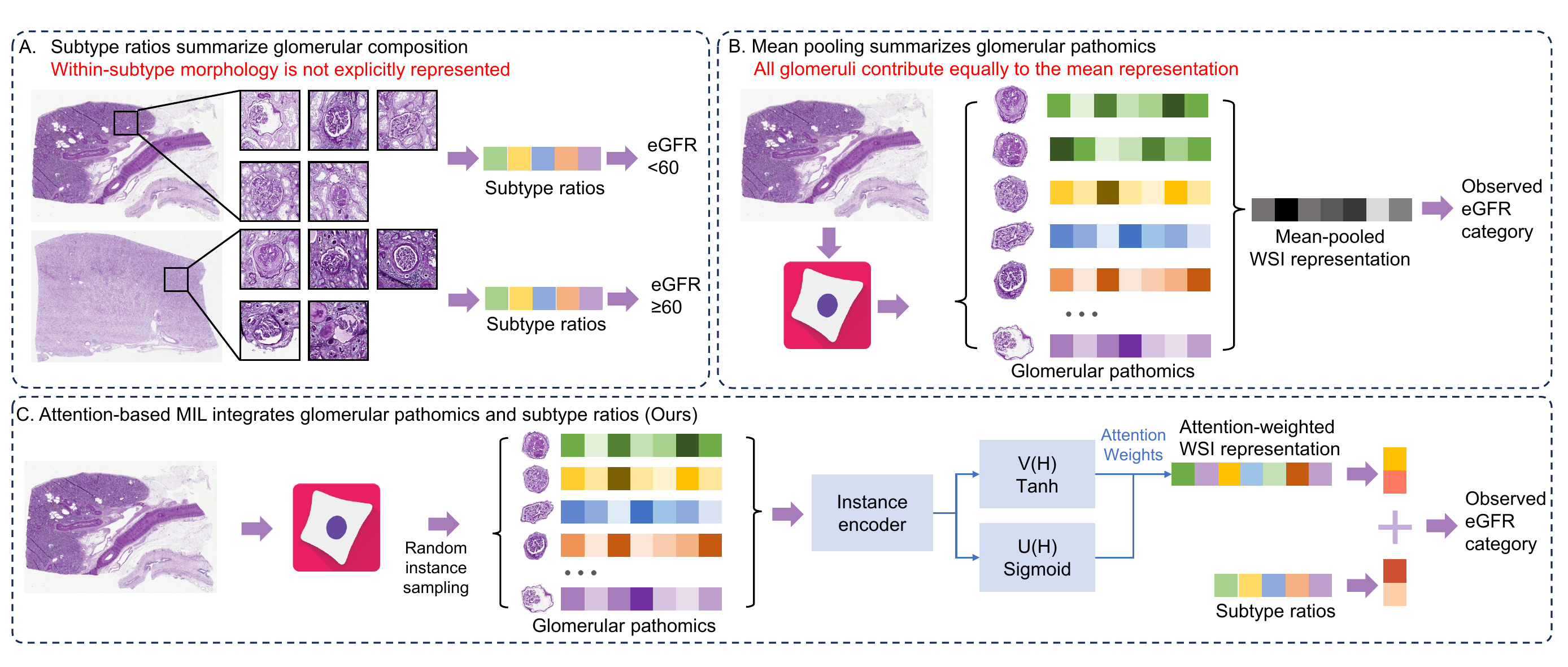}
  \caption{Schematic overview of the three experimental settings evaluated for eGFR classification. \textbf{(A) Subtype Ratios:} Uses only the global proportions of glomerular subtypes (e.g., viable, ischemic, globally sclerotic) as the patient-level feature. \textbf{(B) Mean Pooling:} A baseline architecture that extracts pathomics features from individual glomeruli and aggregates them via uniform mean pooling to generate a patient-level representation, treating all instances equally. \textbf{(C) Attention-based MIL:} The proposed pipeline, which performs random instance sampling ($n_s=256$), instance encoding, and gated attention pooling to generate a weighted pathomics representation. This representation is late-fused with the glomerular subtype ratios to yield the final eGFR prediction. In panel (C), the fusion symbol represents the learnable logit-level combination defined in Eq.~\ref{eq:logit_fusion}.}
  \label{fig:architecture}
\end{figure}

Chronic kidney disease (CKD) is a major global health burden associated with progressive loss of renal function, increased cardiovascular risk, and substantial healthcare costs~\cite{huo2021ai,mark2025global}. Accurate assessment of chronic kidney injury is therefore important for diagnosis, prognosis, and treatment planning. In routine renal pathology, whole-slide images (WSIs) of kidney biopsies provide important histological evidence for evaluating disease severity. Glomeruli are particularly informative because their structural alterations are associated with the development and progression of many renal diseases~\cite{yang2020circlenet,jha2021instance,kim2026glomerulosclerosis}. In quantitative renal pathology, the proportions of different glomerular subtypes, particularly the percentage of globally sclerotic glomeruli, are well-established measures of chronic renal injury and have demonstrated clinical and prognostic value in both native and transplant kidney disease~\cite{deng2026kpis,denic2022prognostic,denic2022changes}. These subtype ratios provide a compact and interpretable summary of the glomerular composition of a biopsy. In addition to subtype composition, individual glomeruli exhibit quantitative variations in size, shape, staining characteristics, texture, and structural organization~\cite{yu2025glo,denic2022prognostic}. These variations may reflect differences among glomeruli assigned to the same pathological subtype and could provide information complementary to established subtype ratios. Preserving and analyzing this glomerulus-level heterogeneity may therefore provide a more comprehensive quantitative characterization of renal injury~\cite{guo2026glo,lu2021improve,denic2022changes}.

Recent advances in computational pathology have enabled the systematic extraction of quantitative features, commonly referred to as pathomics~\cite{chen2024spatial,yang2025pyspatial}, from segmented histopathological structures~\cite{deng2023omni,deng2024prpseg,zhao2026mori,deng2024hats}. Pathomic analysis has shown potential for disease classification, outcome prediction, and biomarker discovery by quantifying tissue characteristics that may not be fully represented by categorical labels~\cite{yin2025differential}. A common implementation summarizes instance-level pathomic features using slide-level statistics, such as their mean or variance. Although these representations are concise and computationally efficient, mean pooling assigns equal importance to all glomeruli and may reduce subtle information from less prevalent but potentially informative glomerular patterns. Multiple-instance learning (MIL) provides an alternative framework in which each patient is represented as a bag of individual glomerular instances while requiring only a patient-level label~\cite{reisenbuchler2025top,deng2024cross}. In computational pathology, MIL has been widely applied using image patches or learned image embeddings as instances~\cite{zhang2025patches,deng2024cross}. However, relatively few studies have investigated MIL using handcrafted pathomic features extracted from biologically defined structures such as glomeruli. Attention-based multiple-instance learning (ABMIL) can learn the relative importance of individual glomeruli while preserving the quantitative meaning of the original pathomic features~\cite{ilse2018attention,kurata2025multiple,kasireddy2025automated}. Whether ABMIL-based glomerular pathomics can provide additional information for kidney-biopsy analysis, particularly in estimated glomerular filtration rate (eGFR) classification, remains insufficiently investigated.

In this proof-of-concept study, we develop an interpretable ABMIL framework for eGFR classification using glomerulus-level pathomic features extracted from kidney-biopsy WSIs. Individual glomeruli are first identified and classified into pathological subtypes, after which quantitative pathomic features are extracted from each glomerulus. Instead of representing these features only through slide-level averages, the proposed framework treats each glomerulus as an instance and learns to aggregate heterogeneous glomerular profiles into a patient-level representation. We compare ABMIL with a multilayer perceptron (MLP) using mean-pooled pathomic features. The MLP is evaluated using three input settings: glomerular subtype ratios only, pathomics only, and their combination. ABMIL is evaluated using pathomics alone and pathomics combined with subtype ratios. SHapley Additive exPlanations (SHAP) are further used to examine the contributions of individual quantitative features, as depicted in the schematic overview of the experimental settings (Fig.~\ref{fig:architecture}). Through these experiments, we investigate whether MIL-based glomerular pathomics provides information complementary to established subtype ratios for quantitative kidney-biopsy assessment.

\section{Method}

\begin{figure}[htbp]
  \centering
  \includegraphics[width=1.0\textwidth]{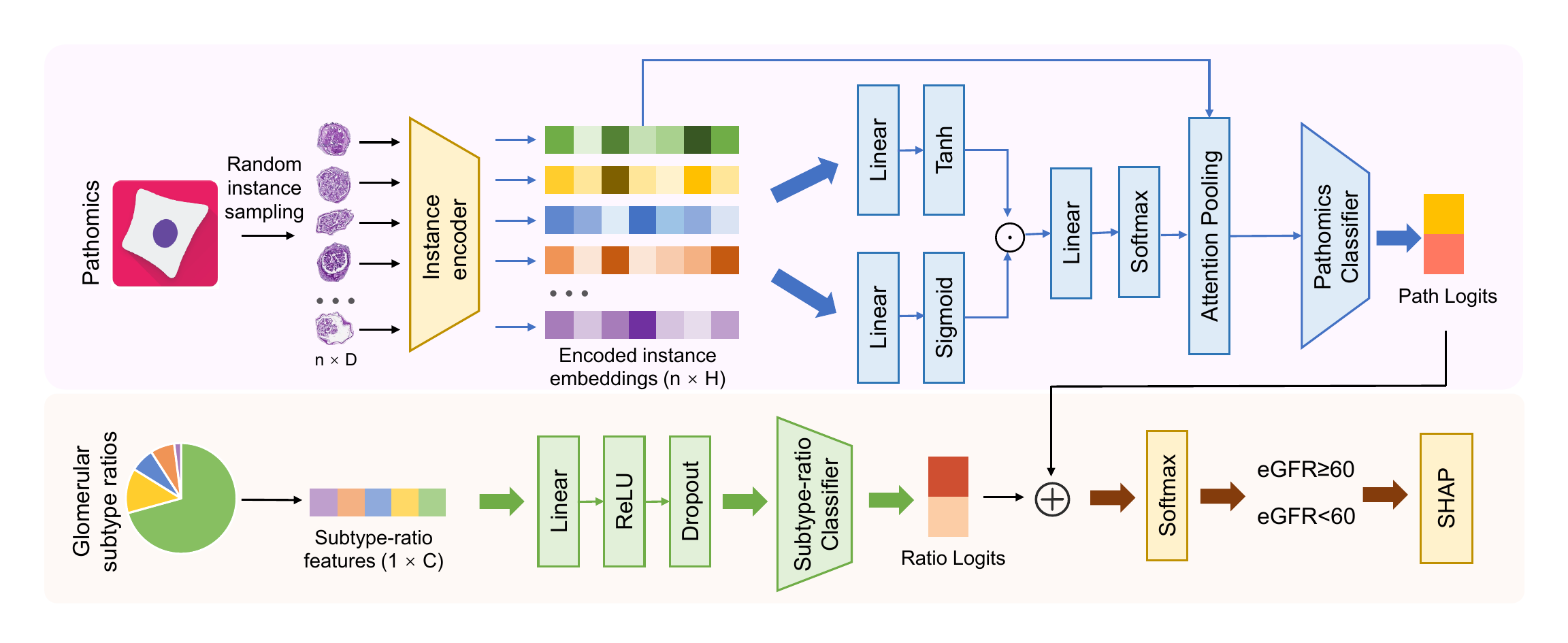}
  \caption{Overview of the proposed pipeline for predicting eGFR categories. The framework comprises two parallel branches. The \textbf{Pathomics branch} first performs random instance sampling ($n_s=256$) for each patient, followed by an instance encoder and gated attention pooling to aggregate instance-level embeddings into a patient-level representation. This representation is passed through a dedicated classifier to produce Pathomics Logits. The \textbf{Glomerular subtype ratios branch} processes the global subtype ratio features through a multi-layer perceptron (Linear, ReLU, Dropout) to generate Ratio Logits, which are then combined with the Pathomics Logits via a learnable convex combination. The fused logits are then passed through a softmax function to yield the final eGFR classification ($\geq 60$ vs. $< 60$). The SHAP block denotes a separate post hoc attribution analysis of the fixed-dimensional attention-aggregated representation. It is not part of the prediction pathway and should not be interpreted as an exact decomposition of the complete ABMIL bag-level decision function.}
  \label{fig:workflow}
\end{figure}

We propose an interpretable ABMIL framework for classifying patients into $\mathrm{eGFR}<60~\mathrm{mL/min/1.73\,m^2}$ and $\mathrm{eGFR}\geq60~\mathrm{mL/min/1.73\,m^2}$ categories from kidney-biopsy WSIs, as illustrated in Fig.~\ref{fig:workflow}. The framework models each patient using glomerulus-level pathomic features and biopsy-level glomerular subtype ratios. ABMIL aggregates heterogeneous glomerular instances into a patient-level pathomics representation, which is integrated with the glomerular subtype ratios through learnable logit-level fusion. SHAP is subsequently used for post hoc feature attribution.

\subsection{Glomerular Pathomic Representation}
\label{subsec:pathomics}

Individual glomeruli were first detected, segmented, and classified into pathological subtypes using an automated glomerular analysis framework~\cite{shi2026m}. Within our cohort of 162 whole-slide images, 17 slides were manually annotated by an expert pathologist to establish the ground truth for training the framework, while the glomerular structures on the remaining 145 slides were inferred automatically by the model without subsequent manual quality control. Quantitative pathomic features were then extracted from each segmented glomerulus using a custom Python pipeline following feature definitions implemented in CellProfiler~\cite{stirling2021cellprofiler}. The extracted features characterized glomerular size, shape, staining intensity, texture, and radial intensity distribution.

For a patient containing $n$ detected glomeruli, the pathomic input is represented as a bag $\mathcal{X}=\{\mathbf{x}_1,\ldots,\mathbf{x}_n\}$, where $\mathbf{x}_i\in\mathbb{R}^{D}$ denotes the $D$-dimensional pathomic feature vector of glomerulus $i$. This representation preserves glomerulus-level variation rather than reducing all pathomic features to slide-level means.

\subsection{Attention-Based Multiple-Instance Learning}
\label{subsec:abmil}

Each glomerular feature vector is transformed into a latent embedding $\mathbf{h}_i=f_{\mathrm{enc}}(\mathbf{x}_i)$ by an instance encoder. A gated-attention mechanism assigns a normalized attention weight to each glomerulus and aggregates the instance embeddings into a patient-level pathomics representation:
\begin{equation}
\begin{aligned}
s_i
&=
\frac{1}{\tau}
\mathbf{w}^{\mathsf{T}}
\left[
\tanh(\mathbf{V}\mathbf{h}_i)
\odot
\sigma(\mathbf{U}\mathbf{h}_i)
\right],
\\
a_i
&=
\frac{\exp(s_i)}
{\sum_{j=1}^{n}\exp(s_j)},
\\
\mathbf{z}_{\mathrm{path}}
&=
\sum_{i=1}^{n}a_i\mathbf{h}_i,
\end{aligned}
\label{eq:abmil_aggregation}
\end{equation}
where $\odot$ denotes element-wise multiplication, $\sigma(\cdot)$ is the sigmoid function, and $\tau$ is a learnable temperature parameter. The aggregation in Eq.~\ref{eq:abmil_aggregation} allows individual glomeruli to contribute differently to the patient-level representation.

\subsection{Logit-Level Fusion with Glomerular Subtype Ratios}
\label{subsec:fusion}

Let $\mathbf{r}\in\mathbb{R}^{C}$ denote the glomerular subtype-ratio vector, where each component represents the proportion of glomeruli assigned to one of $C$ pathological subtypes. Specifically, $C=5$ subtypes are included: global glomerulosclerosis (GGS), viable glomerulus (Viable Glom), ischemic glomerulus (Ischemic Glom), segmental glomerulosclerosis (SGS), and atubular glomerulus (Atubular Glom). Each ratio is calculated as the count of glomeruli in a specific subtype divided by the total number of detected glomeruli for that patient. Consequently, the sum of all components in $\mathbf{r}$ equals 1. The attention-pooled pathomics representation and glomerular subtype-ratio vector are processed using separate classifiers. Their outputs are integrated through a learnable convex combination:
\begin{equation}
\begin{aligned}
\boldsymbol{\ell}_{\mathrm{path}} &= f_{\mathrm{path}}(\mathbf{z}_{\mathrm{path}}), \quad \boldsymbol{\ell}_{\mathrm{ratio}} = f_{\mathrm{ratio}}(\mathbf{r}), \quad \alpha = \sigma(\beta), \\
\hat{\mathbf{p}} &= \operatorname{softmax} \left[ \alpha\boldsymbol{\ell}_{\mathrm{path}} + (1-\alpha)\boldsymbol{\ell}_{\mathrm{ratio}} \right].
\end{aligned}
\label{eq:logit_fusion}
\end{equation}
In Eq.~\ref{eq:logit_fusion}, $f_{\mathrm{path}}(\cdot)$ and $f_{\mathrm{ratio}}(\cdot)$ denote the pathomics and subtype-ratio classifiers, respectively. $\beta$ is a learnable scalar parameter, and $\alpha=\sigma(\beta)\in(0,1)$ controls their relative contributions, initialized such that $\alpha=0.5$ (i.e., $\beta=0$) to ensure both branches contribute equally at the start of training. The predicted eGFR-category probabilities are denoted by $\hat{\mathbf{p}}$.

\subsection{SHAP-Based Feature Attribution}
\label{subsec:shap}

KernelSHAP~\cite{lundberg2017unified} was used to examine the contributions of individual pathomic features and glomerular subtype ratios to eGFR classification. Directly applying KernelSHAP to the complete bag of instance-level features would result in a high-dimensional attribution problem. We therefore constructed a fixed-length representation by applying the learned attention weights to the original pathomic features and combining the resulting vector with the glomerular subtype ratios:
\begin{equation}
\widetilde{\mathbf{x}}
=
\sum_{i=1}^{n}a_i\mathbf{x}_i,
\qquad
\mathbf{u}
=
[\widetilde{\mathbf{x}};\mathbf{r}],
\label{eq:shap_representation}
\end{equation}
where $[\cdot;\cdot]$ denotes concatenation. KernelSHAP perturbs the components of the fixed-length representation defined in Eq.~\ref{eq:shap_representation} to estimate post hoc feature attributions. Specifically, the attention weights $a_i$ used to construct $\widetilde{\mathbf{x}}$ are pre-computed and held fixed during the perturbation process. For each perturbed sample, the aggregated pathomic portion is treated as a single-instance bag and routed through the trained ABMIL pipeline (i.e., the instance encoder, attention pooling, and pathomics classifier), late-fused with the subtype-ratio branch, and finally passed through a softmax function. Consequently, the SHAP values explain the final predicted class probabilities ($\hat{\mathbf{p}}$) rather than the intermediate logits.

Because the attribution analysis operates on an attention-aggregated representation, the resulting SHAP values should be interpreted as feature attributions for the derived patient-level representation rather than an exact decomposition of the complete ABMIL bag-level decision function.

\section{Data \& Experiments}  

\subsection{Data}
\label{subsec:data}

This study retrospectively included kidney tissue sections derived from non-cancerous parenchyma of nephrectomy cases processed at the Department of Pathology at Northwell Health (comprising specimens from Long Island Jewish Medical Center and Northshore University Hospital) between July 2017 and October 2019. The de-identified tissue sections were stained with periodic acid–Schiff (PAS) and digitized using a Leica GT450 RUO scanner at 20$\times$ magnification. The study was approved by the Institutional Review Board (IRB) of Northwell Health.

Initially, 196 whole-slide images (WSIs) were screened. Thirty-four images were excluded due to missing eGFR clinical values or to ensure a fair comparison where only one WSI per patient was included, resulting in a final cohort of 162 WSIs from 162 patients. Renal function was categorized as $\mathrm{eGFR}<60~\mathrm{mL/min/1.73\,m^2}$ or $\mathrm{eGFR}\geq60~\mathrm{mL/min/1.73\,m^2}$, with 74 and 88 patients in the respective groups. 

Glomeruli within each WSI were detected, segmented, and classified into pathological subtypes using a previously developed automated framework~\cite{shi2026m}. Within our cohort, 17 slides were manually annotated by expert renal pathologists in QuPath using pixel-level instance masks to establish the ground truth for training the framework, while the glomerular structures on the remaining 145 slides were inferred automatically by the model without subsequent manual quality control. In total, 58,290 glomeruli were detected (median 333 glomeruli per patient). 

Glomeruli were classified into five clinically significant pathological subtypes: GGS, Viable Glom, Ischemic Glom, SGS, and Atubular Glom. The patient-level ratio of each subtype was calculated as its number of glomeruli divided by the total number of detected glomeruli.

\subsection{Experimental Details}
\label{subsec:experimental_details}

\noindent\textbf{Pathomic feature preprocessing.}
Features with more than 50\% missing values or fewer than two unique values were removed. The retained features were converted to numeric form, imputed using training-fold medians, and standardized using a \texttt{RobustScaler} fitted exclusively on the training folds to prevent data leakage. No additional stain or intensity normalization was applied. The final feature dimensionality was $D = 308$.

\noindent\textbf{Experimental settings and baselines.}
We evaluated three inputs: glomerular subtype ratios, pathomics, and their combination. The MLP baseline was evaluated with all three inputs. Glomerulus-level pathomic features were mean-pooled for the MLP, whereas ABMIL (Ours) directly processed bags of glomerular instances. ABMIL was evaluated with pathomics alone and with both inputs, but not with subtype ratios alone because these ratios are patient-level features. In the combined ABMIL model, separate pathomics and subtype-ratio branches were integrated by the learnable logit-level fusion in Eq.~\ref{eq:logit_fusion}. Performance was evaluated using stratified, patient-level five-fold cross-validation. The five resulting configurations are reported in Table~\ref{tab:egfr_results}.

\noindent\textbf{Model configuration and training.}

For ABMIL, $n_s=256$ glomeruli were randomly sampled with replacement from each patient to obtain fixed-size bags, including for patients with fewer than 256 detected glomeruli. The instance encoder and gated-attention projection had dimensions of 64 and 32, respectively. The logits from five independently sampled voting rounds were averaged before the final softmax operation. Both the pathomics and subtype-ratio branches used two-layer classifiers, and the fusion coefficient was initialized to $\alpha=0.5$. All model configurations were trained using AdamW~\cite{loshchilov2017decoupled} with a learning rate of $10^{-3}$, weight decay of $10^{-2}$, and a batch size of 16. Dropout with a rate of 0.3 was used for regularization. To address class imbalance, focal loss~\cite{lin2017focal} with $\gamma=2.0$ and class weights inversely proportional to the class frequencies in the training folds was used. A fixed random seed of 2027 was used for reproducibility.

\noindent\textbf{Feature attribution.}
KernelSHAP~\cite{lundberg2017unified} was applied to the attention-aggregated representation in Eq.~\ref{eq:shap_representation}. To construct this representation, attention weights and sampled features were averaged across the five voting rounds. The explainer used the aggregated validation-set features as background data, with a sample count limit of $\min(2D + 2048, 4096)$. Mean absolute SHAP values summarized group-level feature importance, and SHAP summary plots showed the magnitude and direction of feature contributions.

\subsection{Evaluation Metrics}
\label{subsec:metrics}

Performance was evaluated using AUC, accuracy (ACC), macro-F1, and weighted-F1. AUC was calculated from the predicted probability of the positive class, defined as $\mathrm{eGFR}\geq60~\mathrm{mL/min/1.73\,m^2}$. To compute ACC and both F1 metrics, class labels were assigned using a standard probability threshold of 0.5. Macro-F1 weights both eGFR categories equally, whereas weighted-F1 weights them by prevalence. All metrics were computed using scikit-learn~\cite{pedregosa2011scikit} and reported to four decimal places as the mean and standard deviation across the five held-out validation folds.

\section{Results}
\label{sec:results}

\subsection{eGFR Classification Performance}
\label{subsec:classification_results}

Table~\ref{tab:egfr_results} summarizes the eGFR classification performance of all five model and input-feature configurations in the study cohort. ABMIL (Ours) with combined pathomics and glomerular subtype ratios achieved the highest AUC, with a mean value of $0.7630 \pm 0.0665$ across the five validation folds.

\begin{table}[htbp]
    \centering
    \caption{eGFR classification performance in the study cohort. Results are reported as mean $\pm$ standard deviation across five held-out validation folds.}
    \label{tab:egfr_results}
    \small
    \begin{adjustbox}{width=0.8\textwidth,center}
    \begin{tabular}{llcccc}
        \toprule
        \textbf{Method} & \textbf{Input features} & \textbf{AUC} & \textbf{ACC} & \textbf{$\mathrm{F1}_{\mathrm{macro}}$} & \textbf{$\mathrm{F1}_{\mathrm{weighted}}$} \\
        \midrule
        \multirow{3}{*}{MLP~\cite{rumelhart1986learning}}
        & Subtype ratios & $0.7262 \pm 0.0596$ & $0.6481 \pm 0.0567$ & $0.6455 \pm 0.0565$ & $0.6461 \pm 0.0553$ \\
        \cmidrule{2-6}
        & Pathomics & $0.6957 \pm 0.0807$ & $0.6231 \pm 0.0521$ & $0.6112 \pm 0.0633$ & $0.6146 \pm 0.0582$ \\
        \cmidrule{2-6}
        & Combined & $0.7162 \pm 0.0331$ & $0.6602 \pm 0.0356$ & \textbf{0.6515} $\pm$ 0.0389 & \textbf{0.6548} $\pm$ 0.0377 \\
        \midrule
        \multirow{2}{*}{ABMIL (Ours)}
        & Pathomics & $0.7362 \pm 0.0788$ & \textbf{0.6667} $\pm$ 0.0976 & $0.6364 \pm 0.1271$ & $0.6461 \pm 0.1182$ \\
        \cmidrule{2-6}
        & Combined & \textbf{0.7630} $\pm$ 0.0665 & $0.6544 \pm 0.0911$ & $0.6500 \pm 0.0948$ & $0.6510 \pm 0.0926$ \\
        \bottomrule
    \end{tabular}
    \end{adjustbox}
\end{table}

Under matched input-feature settings, ABMIL achieved higher mean AUCs than MLP. With pathomics alone, ABMIL achieved an AUC of $0.7362 \pm 0.0788$, compared with $0.6957 \pm 0.0807$ for mean-pooled MLP. With the combined input, ABMIL achieved $0.7630 \pm 0.0665$, compared with $0.7162 \pm 0.0331$ for MLP. These results suggest that attention-based aggregation can retain information from heterogeneous glomerular instances that may be reduced by uniform mean pooling.

Within ABMIL, adding glomerular subtype ratios to pathomics increased the mean AUC from $0.7362$ to $0.7630$. This result supports the primary hypothesis that glomerulus-level pathomics can provide information complementary to established subtype ratios when the two sources are integrated appropriately. However, the combined ABMIL model did not achieve the highest value for every metric. ABMIL with pathomics alone produced the highest ACC, whereas MLP with combined inputs produced the highest macro-F1 and weighted-F1. The relatively large fold-to-fold standard deviations of the ABMIL results also indicate variability across data partitions. Therefore, the principal performance finding is the improvement in AUC rather than uniform superiority across all evaluation metrics.

\begin{figure}[htbp]
  \centering
  \includegraphics[width=1.0\textwidth]{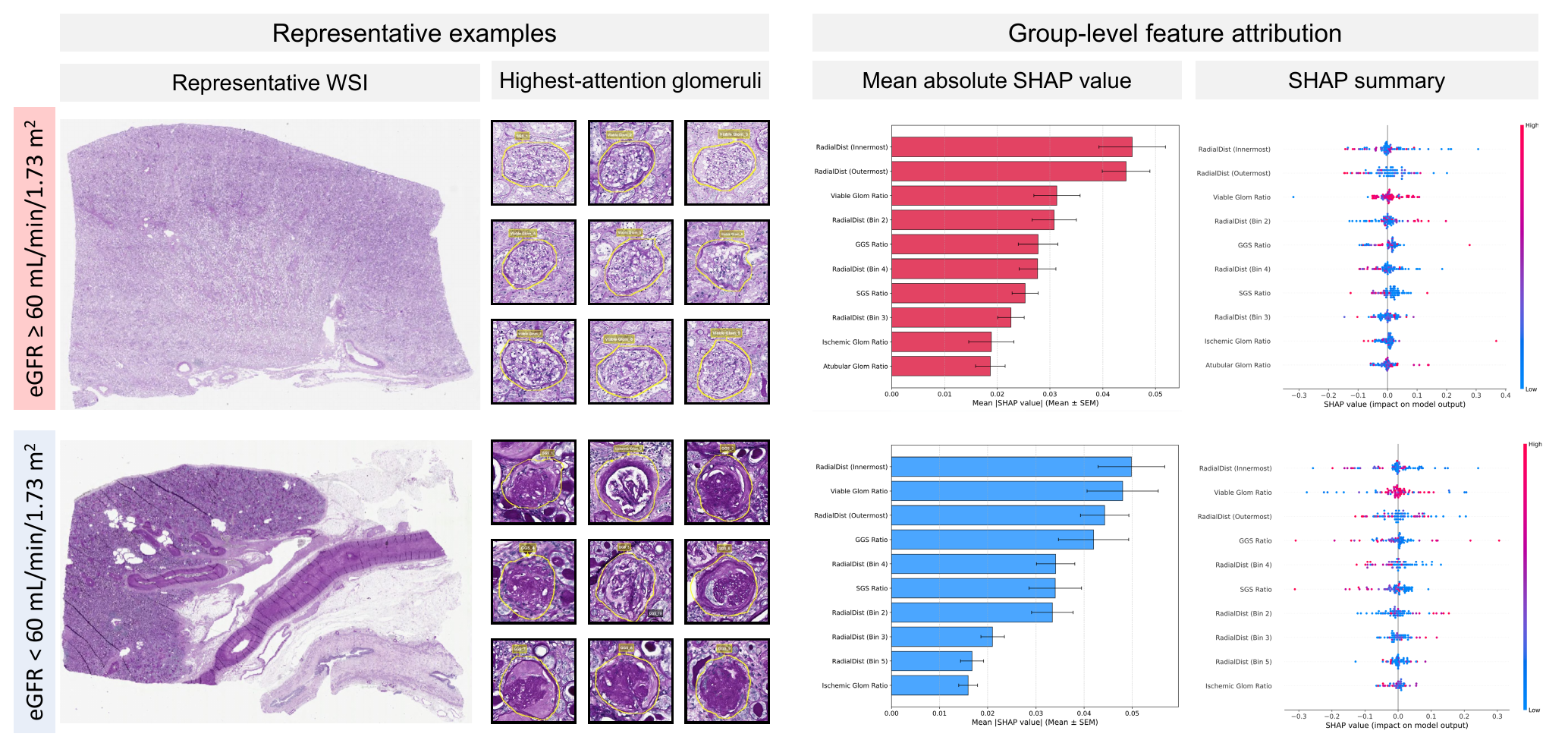}
  \caption{Representative examples and group-level feature attribution for the observed $\mathrm{eGFR}\geq60~\mathrm{mL/min/1.73\,m^2}$ ($n=88$) and $\mathrm{eGFR}<60~\mathrm{mL/min/1.73\,m^2}$ ($n=74$) categories. For each category, the left panel shows a representative WSI, the middle panel shows the nine glomeruli receiving the highest attention weights, and the right panels show the ten highest-ranked features. Bar lengths represent mean absolute SHAP values, and error bars indicate the standard error of the mean (SEM). SHAP summary plots show the distribution and direction of feature contributions across patients, with red and blue indicating relatively high and low feature values, respectively.}
  \label{fig:egfr_interpretation}
\end{figure}

\subsection{Model Interpretation Across eGFR Categories}
\label{subsec:egfr_interpretation}

Fig.~\ref{fig:egfr_interpretation} presents representative WSIs, their highest-attention glomeruli, and group-level feature-attribution results for the observed $\mathrm{eGFR}\geq60~\mathrm{mL/min/1.73\,m^2}$ ($n=88$) and $\mathrm{eGFR}<60~\mathrm{mL/min/1.73\,m^2}$ ($n=74$) categories. 

To ensure consistency, SHAP values for both eGFR groups were computed for the same model output, specifically the predicted probability of the $\mathrm{eGFR}\geq60$ class. The attention examples illustrate the glomerular instances assigned the greatest weights within the WSI shown in the same row. These attention weights were aggregated across the five voting rounds to identify the most salient instances. Meanwhile, the mean absolute SHAP values summarize feature importance within each observed eGFR group.

Radial-distribution features were among the highest-ranked pathomic features in both groups, particularly measurements from the innermost and outermost radial regions of the glomeruli. These features quantify how staining intensity is distributed from the glomerular center toward the periphery, measured across six concentric bins. To enhance readability, code-style feature names in the figure (e.g., \textit{1of6}, \textit{6of6}) have been replaced with descriptive labels corresponding to the innermost and outermost radial bins, respectively. These spatial measurements provide information not represented by subtype ratios alone. The Viable Glom and GGS ratios were also highly ranked, indicating that the attribution analysis retained established measures of glomerular composition alongside instance-level morphological information.

The SHAP summary plots in Fig.~\ref{fig:egfr_interpretation} further show the distribution and direction of individual feature contributions within each observed eGFR category. Although several influential features were shared between the two groups, their attribution distributions were not identical. These results suggest that the model used both glomerular composition and within-glomerulus pathomic variation. The attention and SHAP visualizations should nevertheless be interpreted as descriptive model explanations rather than evidence that the highlighted glomeruli or features have causal relationships with renal function.

\subsection{Feature Attribution Across Global Glomerulosclerosis Strata}
\label{subsec:ggs_attribution}

\begin{figure}[htbp]
  \centering
  \includegraphics[width=0.8\textwidth]{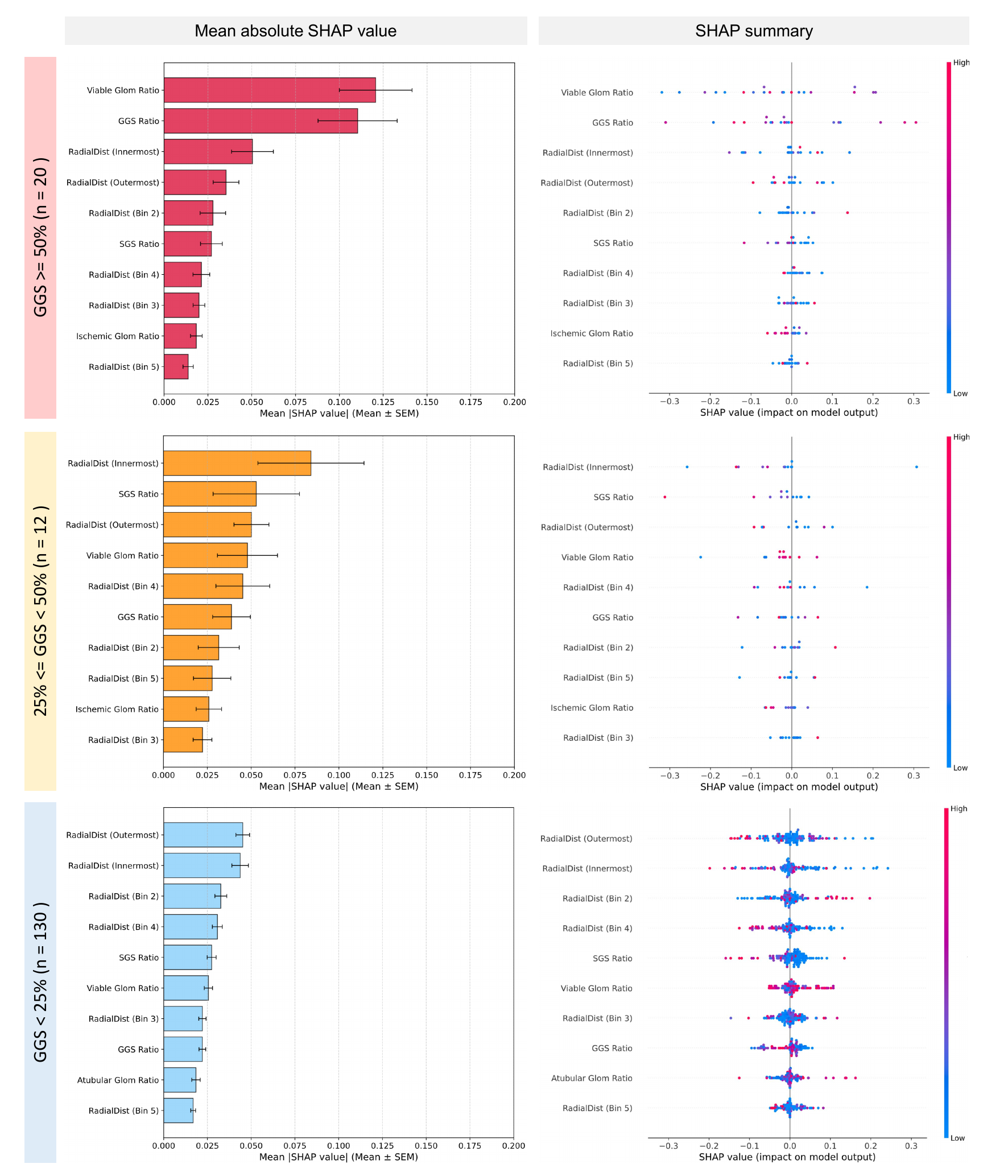}
  \caption{Group-level feature attribution analysis stratified by the percentage of globally sclerotic glomeruli (GGS). Three subgroups are evaluated: GGS $< 25\%$, $25\% \leq$ GGS $< 50\%$, and GGS $\geq 50\%$. For each subgroup, the top 10 features are displayed. \textbf{Left (Bar plots):} Bars represent the mean absolute SHAP values, and error bars indicate the standard error of the mean (SEM), calculated as the standard deviation divided by the square root of the sample size ($n$). \textbf{Right (Scatter plots):} SHAP summary (beeswarm) plots show the distribution and direction of feature contributions across individual patients, where point color indicates the relative feature value (red: high, blue: low). To facilitate direct visual comparison across the GGS strata, the x-axis scales within each plot type (bar and scatter) are held constant across all three subgroups.}
  \label{fig:ggs_attribution}
\end{figure}

Fig.~\ref{fig:ggs_attribution} stratifies the feature-attribution results according to the observed GGS ratio: GGS ratio $\geq50\%$ ($n=20$), $25\%\leq$ GGS ratio $<50\%$ ($n=12$), and GGS ratio $<25\%$ ($n=130$). The same mean absolute SHAP and SHAP summary analyses were used for all three strata.
Consistent with the eGFR analysis, all three strata were evaluated using out-of-fold predictions from the same SHAP procedure, explaining the same model output (the predicted probability of the eGFR $\geq 60$ class).

In the GGS ratio $\geq50\%$ stratum, the Viable Glom and GGS ratios had the largest mean absolute SHAP values, followed by several radial-distribution features. In the intermediate and low GGS-ratio strata, radial-distribution features occupied more of the highest-ranking positions. This pattern suggests that explicit measures of glomerular composition contributed strongly when global glomerulosclerosis was extensive, whereas spatial pathomic measurements remained informative across patients with lower GGS ratios.

The recurrence of radial-distribution features across all three strata supports the complementary role of glomerulus-level morphology. At the same time, these subgroup results are descriptive and should not be interpreted as formal evidence of differences among strata. In particular, the strata were substantially imbalanced, with only 12 patients in the intermediate group and 20 in the high-GGS group. The subgroup attributions therefore require validation in a larger cohort with more balanced GGS-ratio distributions.

\section{New or Breakthrough Work to Be Presented}

This work presents a proof-of-concept framework that formulates glomerulus-level pathomics as an MIL problem and integrates the resulting attention-pooled representation with established glomerular subtype ratios through learnable logit-level fusion. Unlike uniform mean pooling, ABMIL preserves glomerulus-level heterogeneity and learns the relative contributions of individual instances. The framework also supports post hoc feature attribution through a fixed-dimensional, attention-aggregated representation, enabling KernelSHAP analysis of pathomic and subtype-ratio features.

\section{Conclusion}

In this proof-of-concept study, we developed an ABMIL framework for eGFR classification from kidney-biopsy WSIs. The proposed model integrates glomerulus-level pathomics with established glomerular subtype ratios through learnable logit-level fusion. Among the evaluated configurations, ABMIL with combined inputs achieved the highest AUC of $0.7630 \pm 0.0665$, suggesting that instance-level pathomics can provide information complementary to subtype ratios. Attention and KernelSHAP analyses further provided descriptive views of influential glomeruli and quantitative features. These findings demonstrate the feasibility of the proposed approach and motivate validation in larger, independent cohorts.

\section{ACKNOWLEDGMENTS} 
This research was supported by the WCM Radiology AIMI Fellowship, WCM CTSC 2027 Pilot Award, NIH 1R01HL174863-01A1(Sabuncu), and 1U54DK144866-01(Sabuncu). 

\bibliography{main} 
\bibliographystyle{spiebib} 

\end{document}